\documentclass{article}

\usepackage{arxiv}

\usepackage{float}
\usepackage[utf8]{inputenc}
\usepackage[T1]{fontenc}   
\usepackage{hyperref}   
\usepackage{url}      
\usepackage{booktabs}   
\usepackage{amsfonts}
\usepackage{nicefrac}   
\usepackage{microtype} 
\usepackage{lipsum}
\usepackage{amsmath}
\usepackage{graphicx}
\usepackage{subfig}

\usepackage[style=alphabetic,sorting=nyt]{biblatex}
\usepackage{siunitx}
\newcommand{\avgnist}{\ensuremath{\text{avg}_\text{NIST}}}
\newcommand{\E}{\mathbb{E}}
\newcommand{\N}{\mathbb{N}}

\title{Pseudo Random Number Generation through Reinforcement Learning and Recurrent Neural Networks}
\author{
  Luca Pasqualini\thanks{Corresponding author: \url{http://sailab.diism.unisi.it/people/luca-pasqualini/}} \\
  Department of Information Engineering and Mathematical Sciences\\
  University of Siena\\
  Siena, 53100 Italy \\
  \texttt{pasqualini@diism.unisi.it}
  \And
  Maurizio Parton\\
  Department of Economical Studies\\
  University of Chieti-Pescara\\
  Pescara, 65129 Italy\\
  \texttt{parton@unich.it}\\
}

\begin{document}
\maketitle
\begin{abstract}

A Pseudo-Random Number Generator (PRNG) is any algorithm generating a sequence of numbers approximating properties of random numbers.
These numbers are widely employed in mid-level cryptography and in software applications. Test suites are used to evaluate PRNGs quality by checking statistical properties of the generated sequences. These sequences are commonly represented bit by bit.

This paper proposes a Reinforcement Learning (RL) approach to the task of generating PRNGs from scratch by learning a policy to solve a partially observable Markov Decision Process (MDP), where the full state is the period of the generated sequence and the observation at each time step is the last sequence of bits appended to such state.
We use a Long-Short Term Memory (LSTM) architecture to model the temporal relationship between observations at different time steps, by tasking the LSTM memory with the extraction of significant features of the hidden portion of the MDP's states.

We show that modeling a PRNG with a partially observable MDP and an LSTM architecture largely improves the results of the fully observable feedforward RL approach introduced in \cite{pasqualini2020pseudo}.
\end{abstract}

\section{Introduction}\label{sec:introduction}

Generating random numbers is an important task in cryptography, and more generally in computer science. Random numbers are used in several applications, whenever producing an unpredictable result is desirable: for instance games, gambling, encryption algorithms, statistical sampling, computer simulation and modeling, and many others.

An algorithm generating a sequence of numbers approximating properties of random numbers is called a Pseudo-Random Number Generator (PRNG). A sequence generated by a PRNG is ``pseudo-random'' in the sense that it is generated by a deterministic function: the function can be extremely complex, but the same input will give the same sequence. The input of a PRNG is called seed, and to improve the randomness the seed itself can be drawn from a probability distribution. Of course, assuming that drawing from the probability distribution is implemented with an algorithm, everything is still deterministic at a lower level, and the sequence will repeat after a fixed, unknown, number of digits, called \emph{period} of the PRNG.

While true random numbers sequences are more fit to certain applications where true randomness is necessary - for instance, cryptography - they can be very expensive to generate. In most applications, a pseudo-random number sequence is good enough.

The quality of a PRNG is measured by the randomness of the generated sequences. The randomness of a specific sequence can be estimated by running some kind of statistical test suite. In this paper the National Institute of Standards and Technology (NIST) statistical test suite for random and pseudo-random number generators \cite{bassham2010sp} is used to validate the PRNG.

Neural networks have been used for predicting the output of an existing generator, that is, to break the key of a cryptography system. There have also been limited attempts at generating PRNGs using neural networks by exploiting their structure and internal dynamics. For example, the authors of \cite{desai2012using} use recurrent neural networks dynamics to generate pseudo-random numbers. In \cite{hughes2007pseudo}, the authors use the dynamics of a feedforward neural network with random orthogonal weight matrices to generate pseudo-random numbers. Neuronal plasticity is used in \cite{abdi1994neural} instead. In \cite{de2018pseudo} a generative adversarial network approach to the task is presented, exploiting an input source of randomness, like an existing PRNG or a true random number generator.

A PRNG usually generates the sequence incrementally, that is, it starts from the seed at time $t=0$ to generate the first number of the sequence at time $t=1$, then the second at $t=2$, and so on. Thus, it is naturally modeled by a deterministic Markov Decision Process (MDP), where state space, action space and rewards can be chosen in several ways. A Deep Reinforcement Learning (DRL) pipeline can then be used on this MDP to train a PRNG agent. This DRL approach has been used for the first time in \cite{pasqualini2020pseudo} with promising results, see modeling details in Section~\ref{sec:mdp}. This is a probabilistic approach that generates pseudo-random numbers with a ``variable period'', because the learned policy will generally be stochastic. This is a feature of this RL approach.

However, the MDP formulation in \cite{pasqualini2020pseudo} has an action set with size growing linearly with the length of the sequence. This is a severe limiting factor, because when the action set is above a certain size it becomes very difficult, if not impossible, for an agent to explore the action space in a reasonable time.

In this paper, we overcome the above limitation with a different MDP formulation, using a partially observable state. By observing only the last part of the sequence, and using the hidden state of a Long-Short Term Memory (LSTM) neural network to extract important features of the full state, we significantly improve the results in \cite{pasqualini2020pseudo}, see Section~\ref{sec:results}.

The code for this article can be found at GitHub repository \cite{pasqualini2020github}.

\section{Methodology}\label{sec:methodology}

The main idea is quite natural: since a PRNG builds the random sequence incrementally, we model it as an agent in a suitable MDP, and use DRL to train the agent. Several ``hyperparameters'' of this DRL pipeline must be chosen: a good notion of states and actions, a reward such that its maximization gives sequences as close as possible to true random sequences, and a DRL algorithm to train the agent. In this section, we first introduce the DRL notions that will be used in the paper, see Section~\ref{sec:rl}; we then describe the fully observable MDP used in \cite{pasqualini2020pseudo} and the partially observable MDP formulation used in this paper, see Section~\ref{sec:mdp}; in Section~\ref{sec:nist} we describe the reward function used for the MDP; in Section~\ref{sec:nn} we describe the recurrent neural network architecture used to model the environment state; finally, Section~\ref{sec:framework} describe the software framework used for the experiments.

\subsection{Reinforcement Learning (RL)}\label{sec:rl}

For a comprehensive, motivational and thorough introduction to RL, we strongly suggest reading from $1.1$ to $1.6$ in \cite{sutton2018reinforcement}.

RL is learning what to do in order to accumulate as much reinforcement as possible during the course of actions. This very general description, known as \emph{the RL problem}, can be framed as a sequential decision-making problem as follows.

Assume an \emph{agent} is interacting with an \emph{environment}. When the agent is in a certain situation - a \emph{state} - it has several options - called \emph{actions}. After each action, the environment will take the agent to a next state, and will provide it with a numerical \emph{reward}, where the pair "state, reward" may possibly be drawn from a joint probability distribution, called the \emph{model} or the \emph{dynamics} of the environment. The agent will choose actions according to a certain strategy, called \emph{policy} in the RL setting. The RL problem can then be stated as finding a policy maximizing the expected value of the total reward accumulated during the interaction agent-environment.

\begin{figure}[!tbp]
\centering
\includegraphics[width=0.75\textwidth]{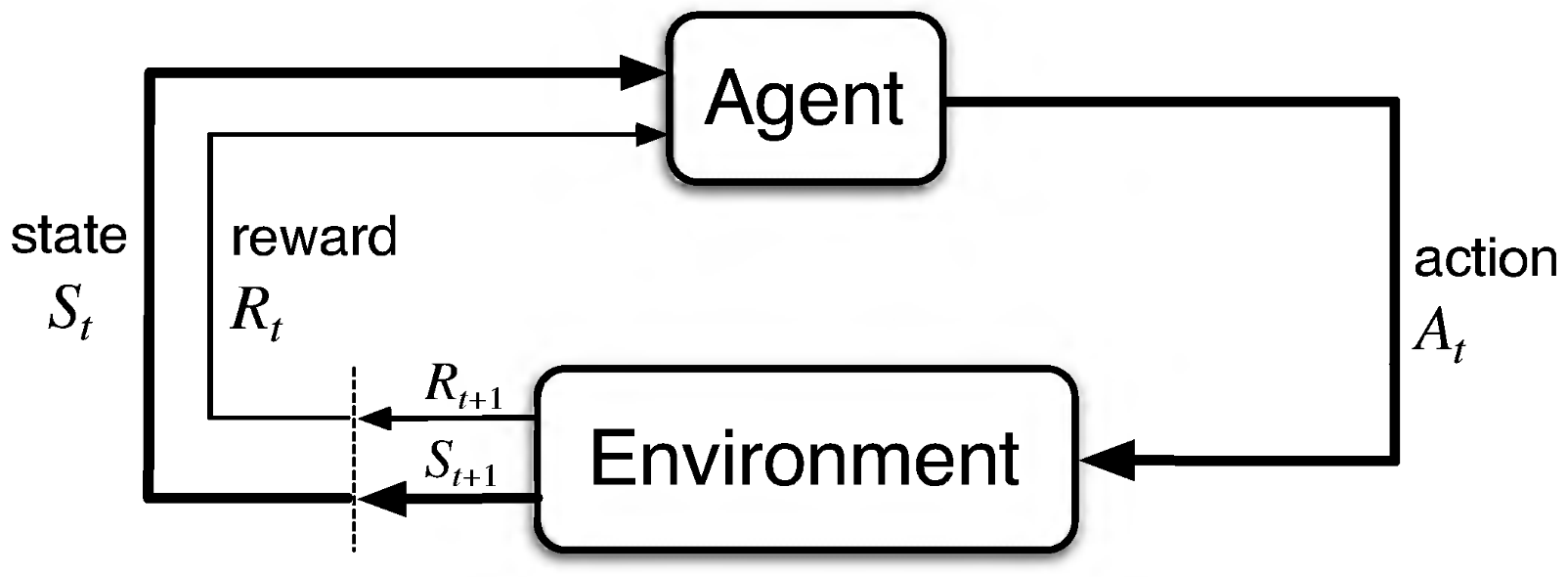}
\caption{\copyright\cite[figure 3.1]{sutton2018reinforcement} The agent-environment interaction is made at discrete time steps $t=0,1,2,\dots$. At each time step $t$, the agent use the state $S_t\in\mathcal{S}$ given by the environment to selects an action $A_t\in\mathcal{A}$. The environment answers with a real number $R_{t+1}\in\mathcal{R}\subset\mathbb{R}$ called reward, and a next state $S_{t+1}$. Going on, we obtain a trajectory $S_0,A_0,R_1,S_1,A_1,R_2,\dots$}\label{fig:mdp}
\end{figure}

To formalize the above description, see Figure~\ref{fig:mdp} representing the agent-environment interaction.

At each time step $t$, the agent receives a state $S_t\in\mathcal{S}$ from the environment, and then selects an action $A_t\in\mathcal{A}$. The environment answers with a numerical reward $R_{t+1}\in\mathcal{R}\subset\mathbb{R}$ and a next state $S_{t+1}$. This interaction gives raise to a \emph{trajectory} of random variables:
\[
S_0,A_0,R_1,S_1,A_1,R_2,\dots
\]
In the case of interest to us, $\mathcal{S},\mathcal{A}$ and $\mathcal{R}$ are finite sets. Thus, the environment answers the action $A_t=a$ executed in the state $S_t=s$ with a pair $R_{t+1}=r,S_{t+1}=s'$ drawn from a discrete probability distribution $p$ on $\mathcal{S}\times\mathcal{R}$, the model (or dynamics) of the environment:
\[
p(s',r|s,a):=p(s',r,s,a):=\text{Pr}(S_{t+1}=s',R_{t+1}=r|S_t=s,A_t=a).
\]
Note the visual clue of the fact that $p(\cdot,\cdot|s,a)$ is a probability distribution for every state-action pair $s,a$.

Figure~\ref{fig:mdp} implicitly assumes that the joint probability distribution of $S_{t+1},R_{t+1}$ depends on the past only via $S_t$ and $A_t$. In fact, the environment is fed only with the last action, and no other data from the history. This means that, for a fixed policy, the corresponding stochastic process $\{S_{t}\}$ is Markov. This gives the name Markov Decision Process (MDP) to the data $(\mathcal{S},\mathcal{A},\mathcal{R},p)$. Moreover, it is a \emph{time-homogeneous} Markov process, because $p$ does not depend on $t$. In certain problems the agent can see only a portion of the full state, called \emph{observation}. In this case, we say that the MDP is \emph{partially observable}. Observations are usually not Markov, because the non observed portion of the state can contain relevant information for the future. In this paper we model a PRNG as an agent in a partially observable MDP.

When the agent experiences a trajectory starting at time $t$, it accumulates a \emph{discounted return} $G_t$:
\[
G_t:=R_{t+1}+\gamma R_{t+2}+\gamma^2 R_{t+3}+\dots=\sum_{k=0}^\infty\gamma^k R_{t+k+1},\qquad\gamma\in[0,1].
\]
The return $G_t$ is a random variable, whose probability distribution depends not only on the model $p$, but also on how the agent chooses actions in a certain state $s$. Choices of actions are encoded by the policy, i.e.\ a discrete probability distribution $\pi$ on $\mathcal{A}$:
\begin{equation*}\label{eq:policy}
\pi(a|s):=\pi(a,s):=\text{Pr}(A_t=a|S_t=s).
\end{equation*}
A discount factor $\gamma<1$ is used mainly when rewards far in the future are less and less reliable or important, or in \emph{continuing} tasks, that is, when the trajectories do not decompose naturally into \emph{episodes}. Since the partially observable MDP formulation we use is episodic, see Section~\ref{sec:mdp}, in this paper we use $\gamma=1$.
 
The average return from a state $s$, that is, the average total reward the agent can accumulate starting from $s$, represents how good is the state $s$ for the agent \emph{following the policy $\pi$}, and it is called \emph{state-value} function:
\[
v_{\pi}(s):=\\E_\pi[G_t|S_t=s].
\]
Likewise, one can define the \emph{action-value} function (known also as \emph{quality} or \emph{q-value}), encoding how good is \emph{choosing an action $a$ from $s$ and then following the policy $\pi$}:
\[
q_{\pi}(s,a):=\\E_\pi[G_t|S_t=s,A_t=a].
\]

Since the return $G_t$ is recursively given by $R_{t+1}+\gamma G_{t+1}$, the RL problem has an optimal substructure, expressed by recursive equations for $v_*$ and $q_*$. If an accurate description of the dynamics $p$ of the environment is available and if one can store in memory all states, then dynamic programming iterative techniques can be used, and an approximate solution $v_*$ or $q_*$ to the Bellman optimality equations can be found. From $v_*$ or $q_*$ one can then easily recover an optimal policy: for instance, $\pi_*(s):=\text{argmax}_{a\in\mathcal{A}}q_*(s,a)$ is a deterministic optimal policy.

However, in most problems we have only a \emph{partial knowledge of the dynamics}, if any. This can be overcome by \emph{sampling trajectories} $S_t=s,A_t=a,R_{t+1},S_{t+1},A_{t+1},R_{t+2},\dots$ to \emph{estimate} the $q$-value $q_\pi(s,a)=\E_\pi[G_t|S_t=s,A_t=a]$, instead of computing a true expectation. Moreover, in most problems \emph{there are way too many states to store them in memory}, or just to go through every state just once. In this case, the estimate of $q_\pi(s,a)$ must be stored in a parametric function approximator $q_\pi(s,a;w)$, where $w$ is a parameters vector living in a dimension much lower than $|\mathcal{S}\times\mathcal{A}|$. Due to their high representational power, deep neural networks are nowadays widely used as approximators in RL: the combination of deep neural networks with RL is called Deep Reinforcement Learning (DRL).

Iterative dynamic programming techniques can then be approximated, giving a family of algorithms known as \emph{Generalized Policy Iteration} algorithms. They work by sampling trajectories to obtain estimates of the true values $\E_\pi[G_t|S_t=s,A_t=a]$, and using supervised learning to find the optimal parameters vector $w$ for $q_\pi(s,a;w)$. This estimated $q$-value is used to find a policy $\pi'$ better than $\pi$, and iterating over this evaluation-improvement loop usually gives an approximate solution to the RL problem.

Generalized Policy Iteration is \emph{value-based}, because uses a value function as a proxy for the optimal policy. A completely different approach to the RL problem is given by Policy Gradient (PG) algorithms. They estimate directly the policy $\pi(a|s;\theta)$, without using a value function. The parameters vector $\theta_t$ at time $t$ is modified to maximize a suitable scalar performance function $J(\theta)$, with the gradient ascent update rule:
\[
\theta_{t+1}:=\theta_t+\alpha\widehat{\nabla J(\theta_t)}.
\]
Here the \emph{learning rate} $\alpha$ is the step size of the gradient ascent algorithm, determining how much we are trying to improve the policy at each update, and $\widehat{\nabla J(\theta_t)}$ is any estimate of the performance gradient $\nabla J(\theta)$ of the policy. Different choices for the estimator corresponds to different PG algorithms. The vanilla choice for the estimator $\widehat{\nabla J(\theta_t)}$ is given by the Policy Gradient Theorem, leading to an algorithm called REINFORCE and to its baselined derivatives, see for instance \cite[Section 13.2 and forward]{sutton2018reinforcement}. Unfortunately, vanilla PG algorithms can be very sensitive to the learning rate, and a single update with a large $\alpha$ can spoil the performance of the policy learned so far. Moreover, the variance of the Monte Carlo estimate is high, and a huge amount of episodes are required for convergence. For this reason, several alternatives for $\widehat{\nabla J(\theta_t)}$ has been researched.

In this paper we use the best performing algorithm in \cite{pasqualini2020pseudo}, a PG algorithm called Proximal Policy Optimization (PPO), described in \cite{schulman2017proximal} and considered state-of-the-art in PG methods. PPO tries to take the biggest possible improvement step on a policy using the data it currently has, without stepping too far and making the performance collapse.

The PPO used in this paper is an instance of PPO-Clip as described by OpenAI at~\cite{openai2018ppo}, with only partially shared value and policy heads. More details on the neural network architecture in Section~\ref{sec:nn}. We use $0.2$ as clip ratio, and early stopping: if the mean KL-divergence of the new policy from the old grows beyond a threshold, training is stopped for the policy, while it continues for the value function. We used a threshold of $1.5\cdot K$, with $K=\num{1e-2}$.

To reduce the variance, we used Generalized Advantage Estimation as in \cite{schulman2015high} to estimate the advantage, with $\gamma = 1$ and $\lambda = 0.95$. Saved rewards $R_t$ were normalized with respect to when they were collected (called rewards-to-go in~\cite{openai2018ppo}).
 
\subsection{Modeling a PRNG as a MDP}\label{sec:mdp}

In \cite{pasqualini2020pseudo} we modeled a PRNG as a fully observable MDP in the following way. Choose a bit length for the sequence, say $B$. The state space is given by all possible bit sequences of length $B$:
\[
\mathcal{S} := \{(b_1, b_2, \dots, b_B): b_n\in\{0,1\}\}.
\]
Action $1_n$ is the action of setting the $n^\text{th}$ bit to $1$, and $0_n$ is the action of setting the $n^\text{th}$ bit to $0$.
The action space is then: 
\[
\mathcal{A} = \bigcup_{n=1}^{B}\{1_n, 0_n\}
\]
This finite MDP formulation has $|\mathcal{S}|=2^B$ and $|\mathcal{A}|=2B$, and was called Binary Formulation (BF) in \cite{pasqualini2020pseudo}.

The main problem with BF is the fact that the size $2B$ of the action set grows linearly with the increasing length $B$ of the sequence. Above a certain size, it's no longer possible to learn a policy with high average score. If the size is big enough, no policy can be learned at all because it's almost impossible for an agent to explore an action space so huge in a reasonable time. Consider that for PRNGs a sequence of $1000$ bits is quite short, while for a RL problem $2000$ actions are way too much.

In this paper we overcome this limitation of BF by hiding a portion of the full pseudo-random sequence, letting the agent see and act only on the last $N$ bits. This removes the correlation between the final length of the sequence and the number of actions, at the cost of introducing a temporal dependency among states, breaking Markovianity and making the resulting MDP partially observable. This new problem is solved by approximating the hidden portion of the state with the hidden state of a recurrent neural network with memory, see Section~\ref{sec:nn} for details. We call this new approach Recurrent Formulation (RF). 

Let $N\in\N$ be the number of bits at the end of the full sequence that we want to expose, or, in other words, let  the observations space be the set $\mathcal{O}:=\{0,1\}^N$. We want the agent to be able to change freely the last $N$ bits, that is, the action space $\mathcal{A}$ coincides with $\mathcal{O}$. We also fix a predetermined temporal horizon $T$ for episodes. This means that the size of the action space is $|\mathcal{A}|=2^N$ for a generated sequence of $T\cdot N$ bits. Both $|\mathcal{A}|$ and length of $S_T$ depend on $N$, but they do not depend on each other. Increasing the length of the generated sequence can then be done by increasing the horizon $T$, leaving constant the action space size.

For example, with $N = 3$, action and observation sets are:
\[
\mathcal{O} = \mathcal{A} = \{[0 0 0]; [0 0 1]; [0 1 0]; [1 0 0]; [0 1 1]; [1 0 1]; [1 1 0]; [1 1 1]\}
\] 
Continuing the example, assume at $t = 0$ we start from a random initial state $S_0 = O_0 = [0 0 1]$, and that $A_0 = [1 1 1]$. Now the full state is $S_1 = [0 0 1 1 1 1]$, but only the last 3 bits $O_1 = [1 1 1]$ are observed by the agent. If the agent now chooses $A_1 = [1 0 1]$, the full state becomes $S_2 = [0 0 1 1 1 1 1 0 1]$, and so on. If we set the episodes length to 100, at the end of the episode the full state is a 300 bits sequence.

Clearly, this formulation can work only if the policy approximator $\pi(\cdot|\cdot;\theta)$ can preserve some information from the time series given by the past observations. Recurrent neural networks can model memory, and for this reason are one of the possible approaches to process time series and, more in general, temporal relationships among data.
This is very useful in RL environments where, from the point of the view of the agent, the Markov property does not hold, as it is typically the case in many partially observable environments. This approach was suggested in~\cite{duell2012solving}, see \cite{wang2018health, chakraborty2019finance} for examples of use cases far different from ours.

State-of-the-art recurrent neural networks for this kind of problems are the Gated Recurrent Unit (GRU) ones, and the Long-Short Term Memory (LSTM) ones. Typically, GRU performs better than LSTM, but for this particular formulation we have experienced good performance with LSTM. Thus, we use LSTM layers to approximate the policy network, see~\ref{sec:nn} for details on the neural network architecture.

\subsection{Reward and NIST Test Suite}\label{sec:nist}

The NIST statistical test suite for random and pseudo-random number generators is the most popular application to test the randomness of sequences of bits. It has been published as a result of a comprehensive theoretical and experimental analysis and may be considered as the state-of-the-art in randomness testing for cryptographic and not cryptographic applications. The test suite has become a standard stage in assessing the outcome of PRNGs shortly after its publication.
 
The NIST test suite is based on statistical hypothesis testing and contains a set of statistical tests specially designed to assess different pseudo-random number sequences properties. Each test computes a test statistic value, function of the input sequence. This value is then used to calculate a P-value that summarizes the strength of the evidence for the sequence to be random. For more details, see~\cite{bassham2010sp}.

If $S_t$ is the sequence produced by the agent at time $t$, the NIST test suite can be used to compute the average P-value of all eligible tests run on $S_t$. Some tests return multiple statistic values: in that case, their average is taken. If a test is failed its value in the average is set to zero. Some tests are not eligible on certain too short sequences, and in this case they are not considered for the average. This average $\avgnist(S_t)$ is used at the end $T$ of each episode as a reward function for the MDP:
\[
R_t =
\begin{cases}
    \avgnist(S_t) & \text{ if } t = T \\
    0               & \text{ otherwise}
\end{cases}
\]
This is the same reward strategy used in \cite{pasqualini2020pseudo}. Note that, since P-values are probabilities, rewards belong to $[0, 1]$, and that NIST test suite accuracy grows with the tested sequence length.

\subsection{Neural Network Architecture}\label{sec:nn}

PPO is an actor-critic algorithm. This means that it requires a neural network for the policy (the actor) and another neural network for the value function (the critic). Moreover, since RF is a partially observable MDP formulation, we need a way to maintain as much information as possible from previous observations, without exponentially increasing the size of the state space. We solve this problem with LSTM layers \cite{hochreiter1997long}.

The neural network used for RF starts with two LSTM layers, with $\text{bias} = 1$ for the forget gate. After the LSTM layers, the network splits in two different subnetworks, one for the policy and one for the value function.
The policy subnetwork has three dense layers with 256, 128 and 64 neurons respectively, stacked. All of them have ReLU activation. After this, a dense layer with $2^N$ neurons provides preferences for the actions, which are turned into probabilities by a softmax activation. This is the policy head.
The value function subnetwork starts exactly as the policy one (but with different weights): three dense layers with 256, 128 and 64 neurons respectively, stacked, all with ReLU activation. At the end, a dense layer with $1$ neuron and no activation for the state value. This is the value head.
All layers have Xavier initialization. For additional details refer to the GitHub repository~\cite{pasqualini2020github}.

BF uses two different neural networks for the policy and the value function.
The policy network has three dense layers with 256, 512 and 256 neurons respectively, stacked. All of them have ReLU activation. After this, the policy head is the same as in RF: a dense layer with $2^N$ neurons and softmax activation.
The same for the value function network: three dense layers with 256, 512 and 256 neurons respectively, stacked, with ReLU activation. After this, the value function head is the same as in RF: a dense layer with $1$ neuron and no activation.
All layers have Xavier initialization. For additional details refer to the GitHub repository~\cite{pasqualini2020github}.

\subsection{Framework}\label{sec:framework}

The framework used for the RL algorithms is USienaRL\footnote{Available on PyPi and also on GitHub: \url{https://github.com/InsaneMonster/USienaRL}.}. This framework allows for environment, agent and interface definition using a preset of configurable models. While agents and environments are direct implementations of what is described in the RL theory, interfaces are specific to this implementation. Under this framework, an interface is a system used to convert environment states to agent observations, and to encode agent actions into the environment. This allows to define agents operating on different spaces while keeping the same environment. By default an interface is defined as pass-through, i.e.\ a fully observable state where agents action have direct effect on the environment. The two fashions of the BF are defined using different interfaces in the implementation. Specifically, the wanderer interface mask out all action resulting in a bit being set to itself, while the baseline uses a simple pass-through interface.

The NIST test battery is run with another framework, called NistRng\footnote{Available on PyPi and also on GitHub: \url{https://github.com/InsaneMonster/NistRng}.}. This framework allows us to easily run a customizable battery of statistical set over a certain sequence. The framework also computes which tests are acceptable over certain sequences, i.e.\ due to their length. Acceptable tests for a certain sequence are called eligible tests. Each test returns a value and a flag stating if the test was successfully exceeded or not by the sequence. If a test is not eligible with respect to a certain sequence it cannot be run and it is skipped.

\section{Results}\label{sec:results}
Our experiments consists on multiple sets of training processes of various BF and RF agents. The goal of these experiments is to measure the performance of the new RF agents and compare it with the results achieved by BF agents.

We consider 3 different agents, all trained by PPO-Clip described in Section~\ref{sec:rl} with an actor-critic neural network described in Section~\ref{sec:nn}.

The agent based on the formulation RF introduced in this paper is called $\pi_{RF}$, and similarly we denote by $\pi_{BF}$ the agent based on BF, that we use as a baseline. We introduce also a third agent, a variation of $\pi_{BF}$ denoted by $\hat{\pi}_{BF}$ and called ``wanderer'': this agent is forced to move as much as possible within the environment by masking out all actions that would keep the agent in the same state. In other words, at time-step $t$ the agent $\hat{\pi}_{BF}$ is forbidden from setting a certain bit to the same value it has at the previous time step $t - 1$.

In the experiments, we optimize a PPO-Clip loss with GAE advantage estimation. Two separated policy loss and value loss are estimated by Adam with mini-batch samples of 32 experiences drawn randomly from a buffer. The buffer is filled by 500 episodes for $\pi_{BF}$ and $\hat{\pi}_{BF}$, and by 1000 episodes for $\pi_{RF}$. Once the buffer is full, a training epoch is performed: thus, for instance, an agent $\pi_{BF}$ building sequences of length $B = 200$ by episodes of length $T = 100$ will start training when the buffer is filled with $500\cdot 100 = 50,000$ experiences, and the epoch will end after $[50,000/32] = 1562$ training steps.
Learning rates are \num{3e-4} for the policy and \num{1e-3} for the value. At the end of the epoch, the buffer is emptied, and the RL pipeline goes on by experiencing new episodes.

We present experimental results in the form of plots. Like in \cite{pasqualini2020pseudo}, the performance metric used is the average total reward across sets of epochs called \emph{volley}. In this paper, a volley is made from two epochs, that is, the plots represent a moving average over a window of two epochs.

\begin{figure}[!tbp]
  \centering
  \subfloat[$\pi_{BF}$ with $B = 80$]{\includegraphics[width=0.5\textwidth]{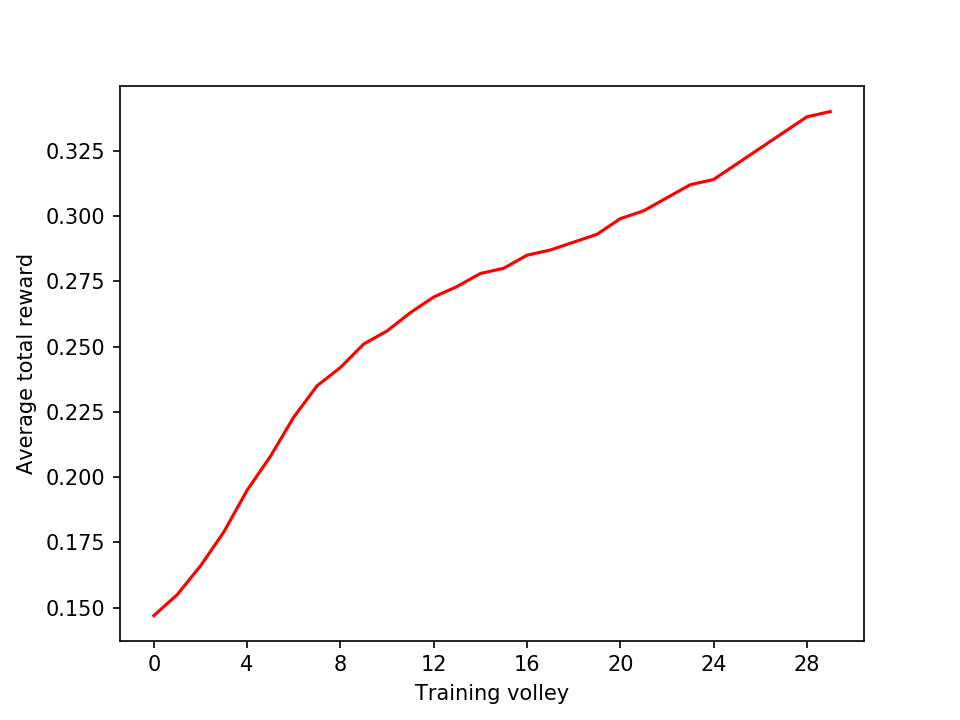}\label{fig:exp_bf_B80_baseline}}
  \hfill
  \subfloat[$\hat{\pi}_{BF}$ with $B = 80$]{\includegraphics[width=0.5\textwidth]{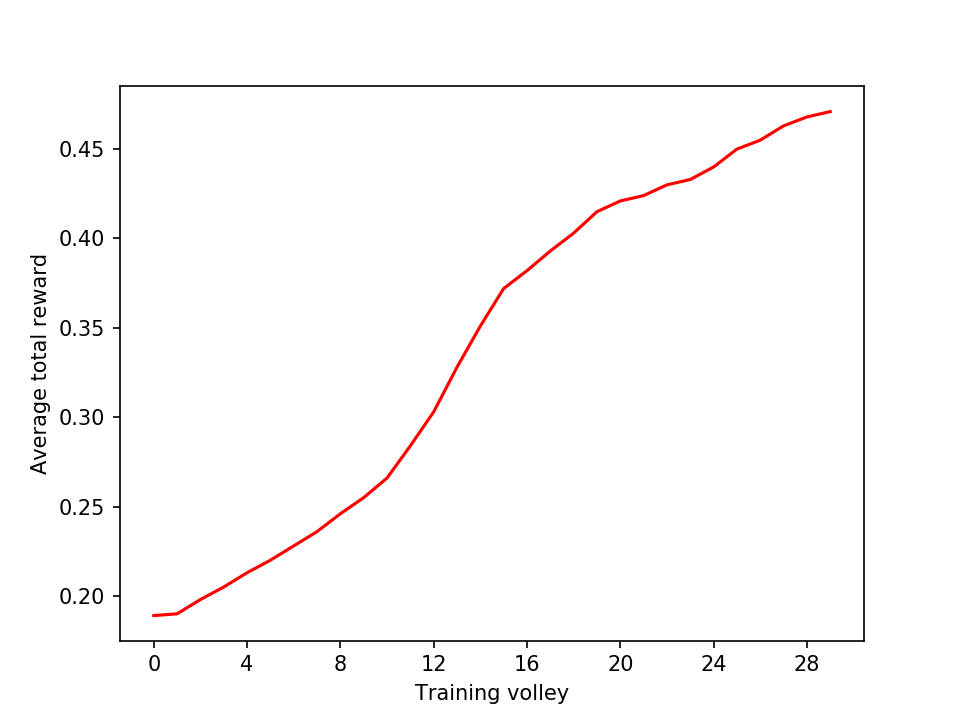}\label{fig:exp_bf_B80_wanderer}}
  \caption{Experiment on BF with $B = 80$. The learning curve is different and the average total reward is better with $\hat{\pi}_{BF}$. Volleys are composed of 1000 episodes each and the fixed length of each trajectory is $T = 40$ steps.}
  \label{fig:BF_exp_N10}
\end{figure}

\begin{figure}[!tbp]
  \centering
  \subfloat[$\pi_{BF}$ with $B = 200$]{\includegraphics[width=0.5\textwidth]{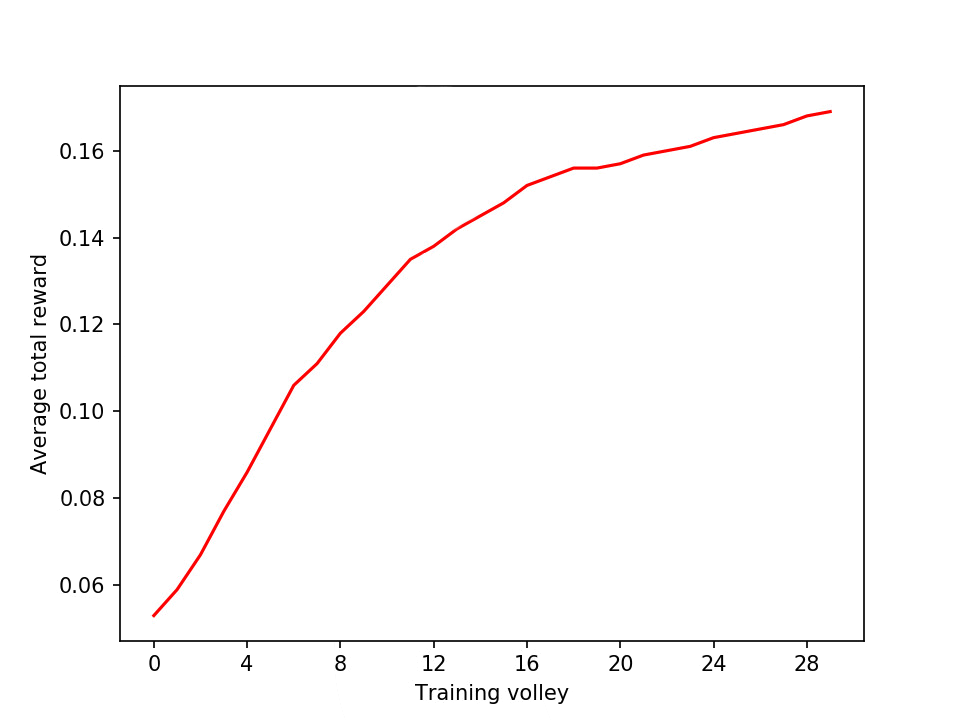}\label{fig:exp_bf_B200_baseline}}
  \hfill
  \subfloat[$\hat{\pi}_{BF}$ with $B = 200$]{\includegraphics[width=0.5\textwidth]{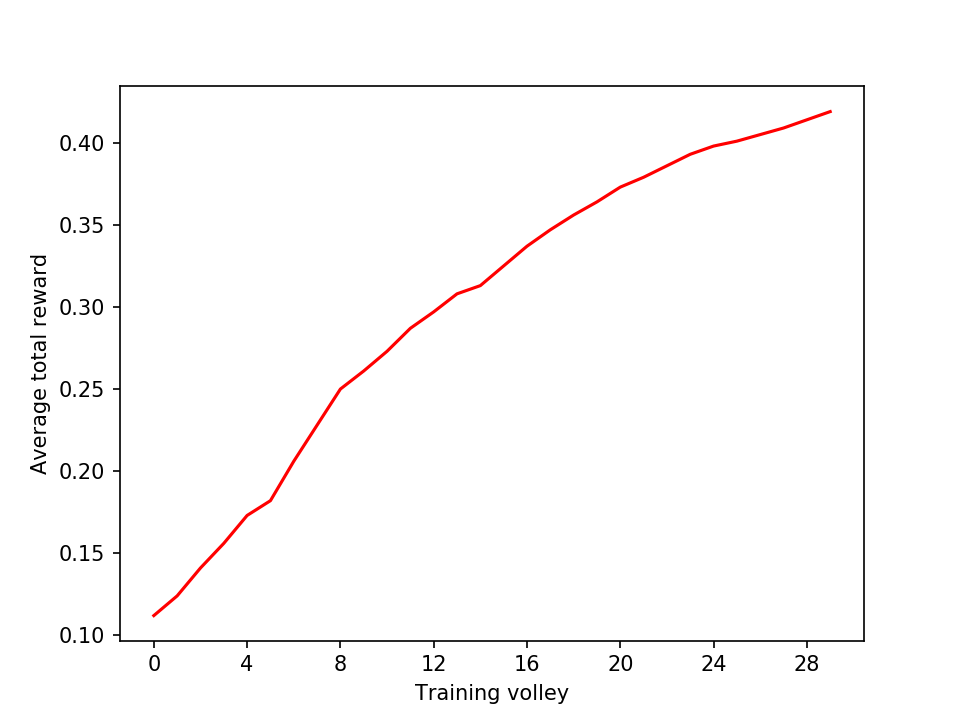}\label{fig:exp_bf_B200_wanderer}}
  \caption{Experiment on BF with $B = 200$. Despite the similar learning curve, there is a huge difference in the achieved average total reward per episode between $\hat{\pi}_{BF}$ and $\pi_{BF}$ at the end of the training process. Volleys are composed of 1000 episodes each and the fixed length of each trajectory is $T = 100$ steps.}
  \label{fig:BF_exp_N25}
\end{figure}

\begin{figure}[!tbp]
  \centering
  \subfloat[$\pi_{BF}$ with $B = 400$]{\includegraphics[width=0.5\textwidth]{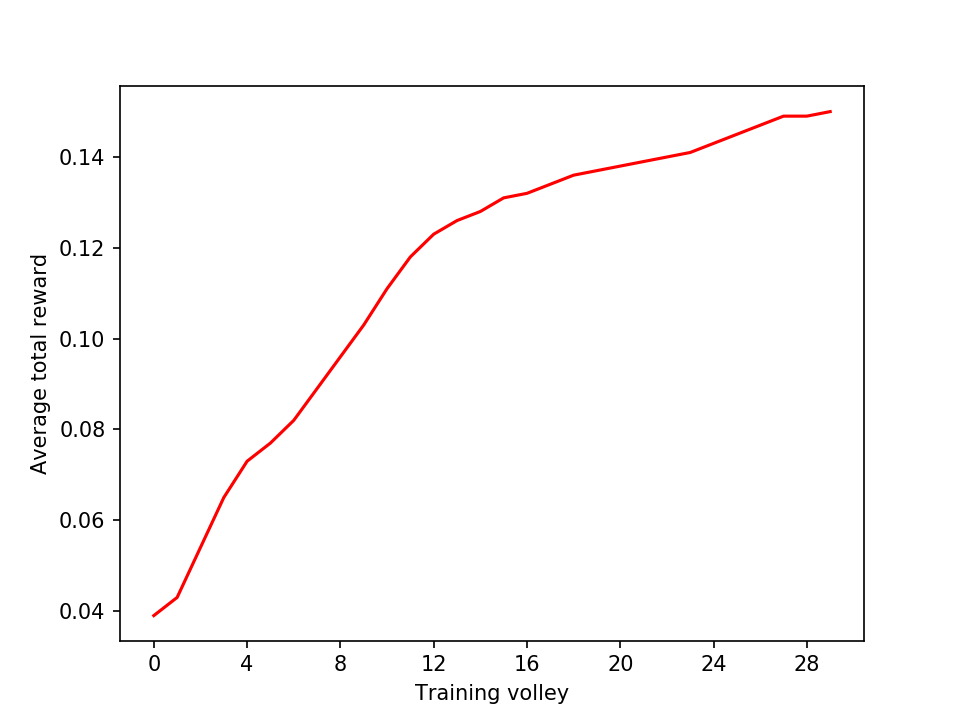}\label{fig:exp_bf_B400_baseline}}
  \hfill
  \subfloat[$\hat{\pi}_{BF}$ with $B = 400$]{\includegraphics[width=0.5\textwidth]{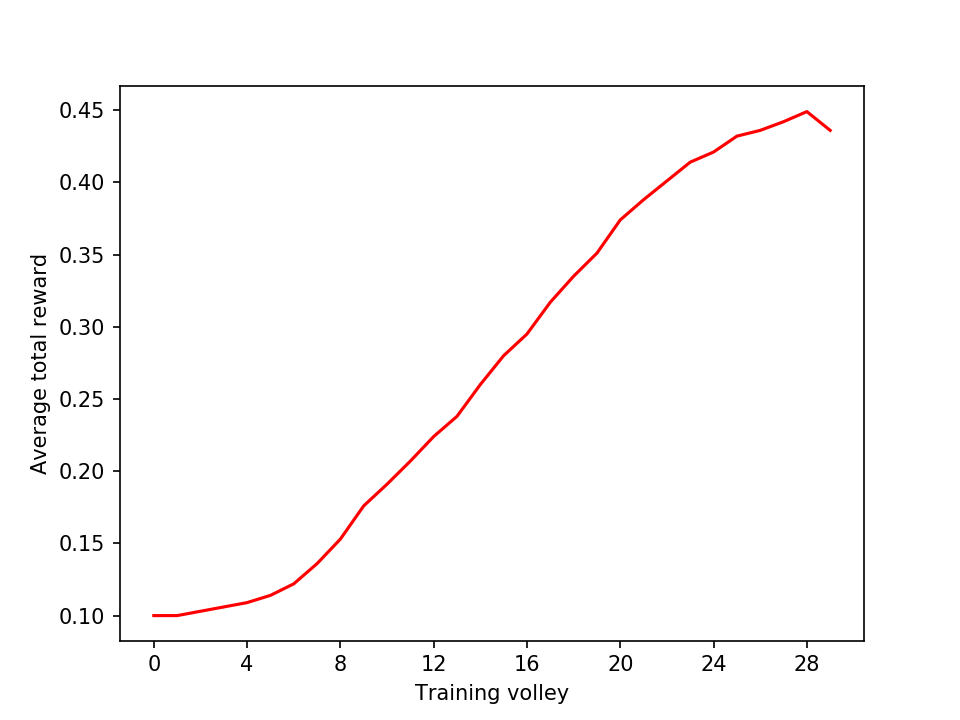}\label{fig:exp_bf_B400_wanderer}}
  \caption{Experiment on BF with $B = 400$. The difference in the achieved average total reward per episode between $\hat{\pi}_{BF}$ and $\pi_{BF}$ is similar to the case with $B = 200$, while the learning curve is different. Volleys are composed of 1000 episodes each and the fixed length of each trajectory is $T = 200$ steps.}
  \label{fig:BF_exp_N50}
\end{figure}

Figures~\ref{fig:exp_bf_B80_baseline}, \ref{fig:exp_bf_B80_wanderer}, \ref{fig:exp_bf_B200_baseline}, \ref{fig:exp_bf_B200_wanderer} and~\ref{fig:exp_bf_B400_baseline}, \ref{fig:exp_bf_B400_wanderer} describe three experiments with $\pi_{BF}$ and $\hat{\pi}_{BF}$ over sequences of different lengths: $B = 80$ bits, $B = 200$ bits and $B = 400$ bits respectively. Length of episodes is $T = 40$, $T = 100$ and $T = 200$ respectively.

For very short sequences of 80 bits, $\hat{\pi}_{BF}$ has better performance at the end yet very similar performance at beginning of training. Making the sequence longer, $B = 200$ bits, allows the wanderer agent to perform much better than the baseline: starting from slightly different performance at the beginning, $\hat{\pi}_{BF}$ has a much higher average total reward at the end of the training process. This trend is confirmed with sequences of $B = 400$ bits.

\begin{figure}
    \centering
    \includegraphics[width=0.65\textwidth]{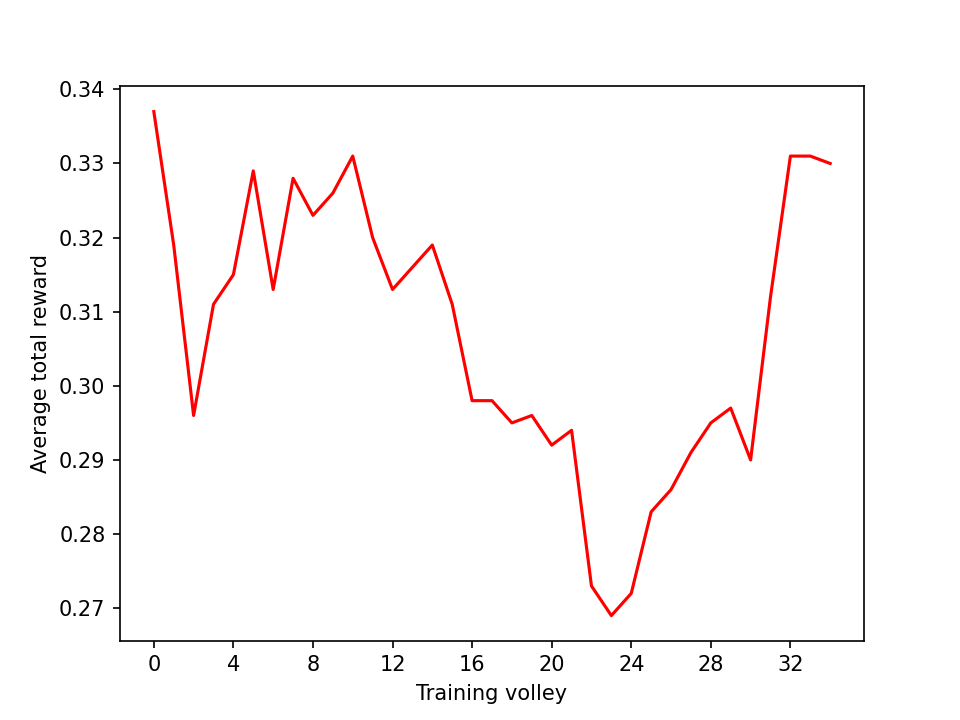}
    \caption{Average total rewards during training of $\pi_{RF}$ with $N = 2$ and $T = 100$. Volleys are composed by 2000 episodes each.}
    \label{fig:exp_rf_N2}
\end{figure}

\begin{figure}
    \centering
    \includegraphics[width=0.65\textwidth]{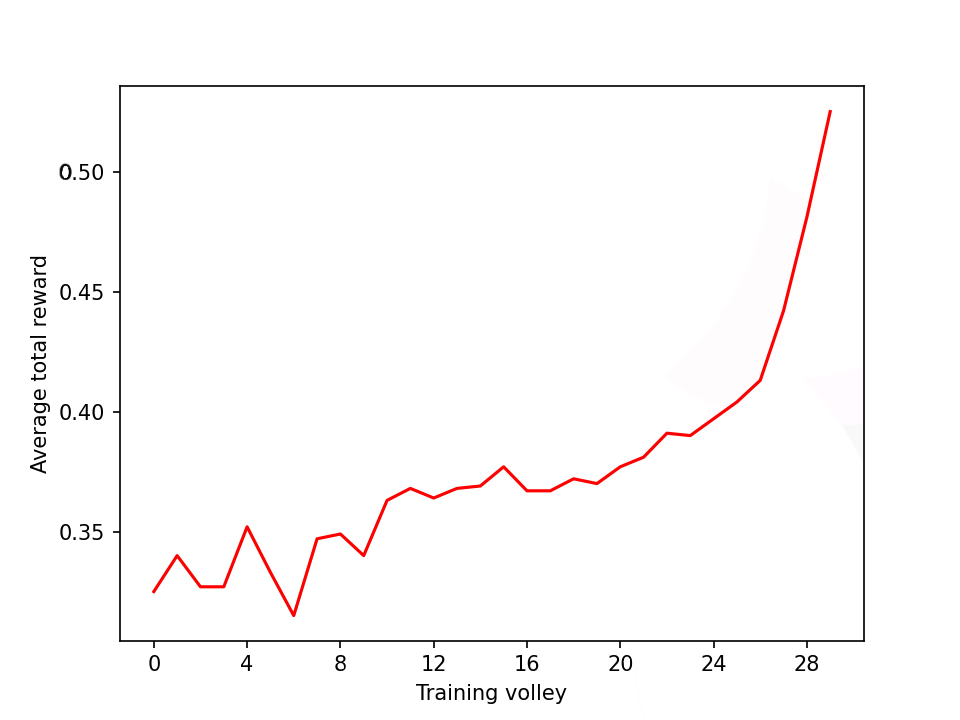}
    \caption{average total rewards during training of $\pi_{RF}$ with $N = 5$ and $T = 100$ steps. Volleys are composed by 2000 episodes each.}
    \label{fig:exp_rf_N5}
\end{figure}

\begin{figure}
    \centering
    \includegraphics[width=0.65\textwidth]{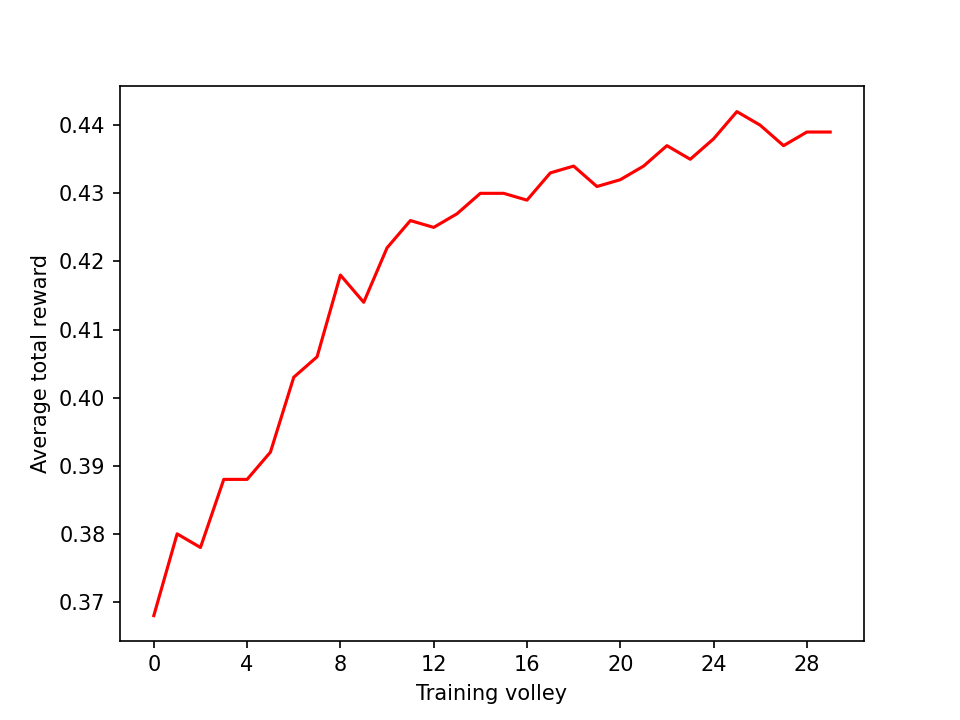}
    \caption{average total rewards during training of $\pi_{RF}$ with $N = 10$ and $T = 100$ steps. Volleys are composed by 2000 episodes each.}
    \label{fig:exp_rf_N10}
\end{figure}

Figures~\ref{fig:exp_rf_N2}, \ref{fig:exp_rf_N5} and \ref{fig:exp_rf_N10} describe three experiments with $\pi_{RF}$. In this case the length of the trajectory $T$ directly influences the length of the sequence generated by $\pi_{RF}$. We keep $T = 100$ constant across $\pi_{RF}$ experiments. This value is chosen experimentally as giving best performance while keeping inference time acceptable.
To obtain sequences comparable with the ones generated by BF agents, we choose to append $N = 2$, $N = 5$ and $N = 10$ bits respectively, resulting in final sequences of length $2 \cdot 100 = 200$ bits, $5 \cdot 100 = 500$ bits and $10 \cdot 100 = 1000$ bits respectively. The seed $S_0$ of the PRNG is drawn from a standard multivariate normal distribution of dimension $N = 2$, $N = 5$ and $N = 10$ respectively.

For $N = 2$, training is not successful. We don't have a clear explanation for this fact. A possible reason is that we are trying to represent the non observable portion of the state with a very low-dimensional input. However, this is not completely true, because in theory the inputs from every time-step is preserved in the hidden state of the LSTM. This issue could be related to the difficulties that recurrent neural networks have shown with gradient descent optimizers, see~\cite{bengio1994learning}.

For $N = 5$, that is, sequences of $500$ bits, $\pi_{RF}$ vastly outperforms $\pi_{BF}$ and, albeit by a narrower margin, $\hat{\pi}_{BF}$ with sequences of $200$ bits. Thus, we have better performance with $150\%$ additional bits in the sequence.
For $N = 10$, that is, sequences of $1000$ bits, $\pi_{RF}$ vastly outperforms $\pi_{BF}$ with $B = 400$. The wanderer agent in this case performs similarly, but still $\pi_{RF}$ produces sequences that are $150\%$ longer than the ones produced by $\hat{\pi}_{BF}$.

We now want to test the hypothesis that, in order to generate PRNGs by DRL, modeling the MDP as in RF is much better than modeling it as in BF.
To this aim, we compare the average total rewards per episode of random agents operating in RF and in the baseline BF.

We call these agents $\rho_{RF}$ and $\rho_{BF}$ respectively.

This comparison is performed to exclude the possibility that PPO is performing better with RF, but maybe different algorithms would perform better with BF. Using a random agent means removing the algorithm from the equation.

\begin{figure}[!tbp]
  \centering
  \subfloat[$\rho_{BF}$ with $B = 80$, episodes length $T = 40$.]{\includegraphics[width=0.5\textwidth]{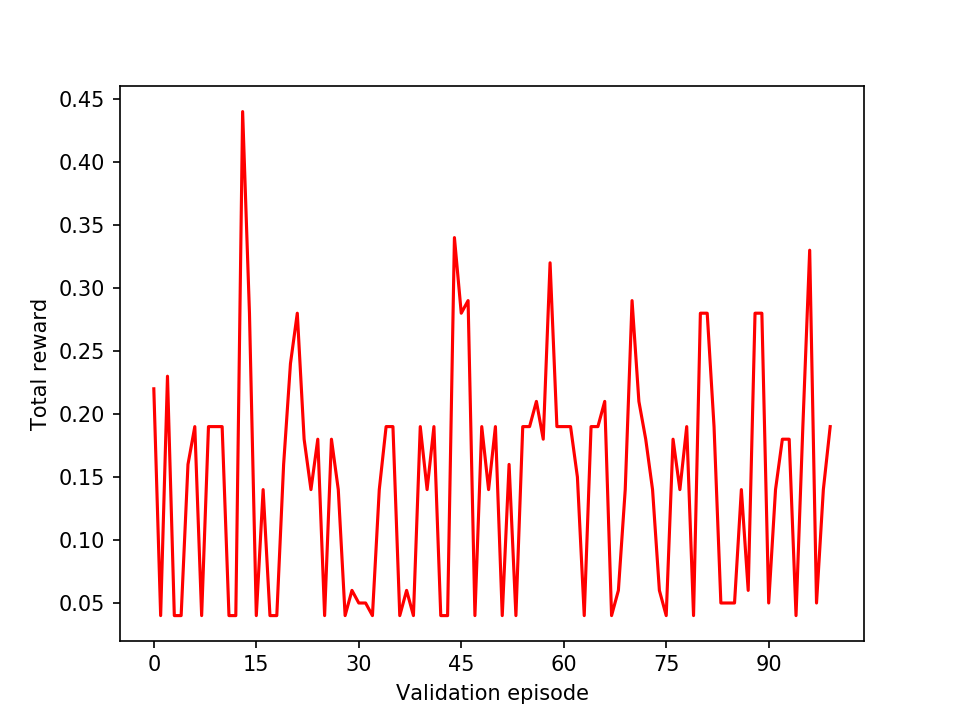}\label{fig:exp_bf_B80_random}}
  \hfill
  \subfloat[$\rho_{RF}$ with $N = 2$, episodes length $T = 100$.]{\includegraphics[width=0.5\textwidth]{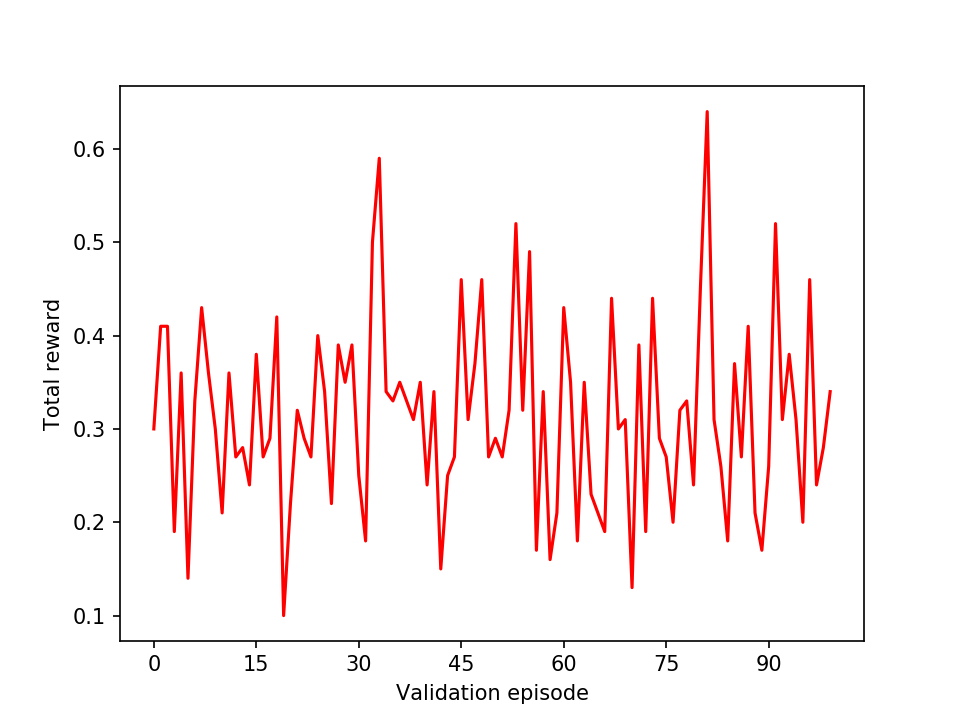}\label{fig:exp_rf_N2_random}}
  \caption{Average total rewards of a random agent on BF and RF for short sequences}
  \label{fig:small_exp_random}
\end{figure}

\begin{figure}[!tbp]
  \centering
  \subfloat[$\rho_{BF}$ with $B = 200$, episodes length $T = 100$.]{\includegraphics[width=0.5\textwidth]{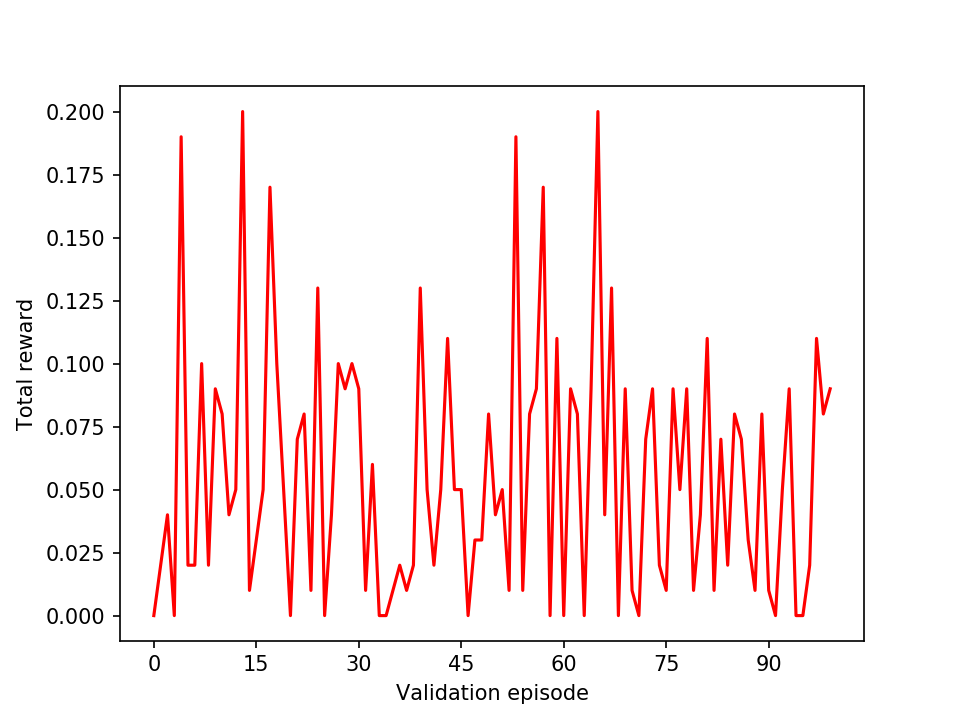}\label{fig:exp_bf_B200_random}}
  \hfill
  \subfloat[$\rho_{RF}$ with $N = 5$, episodes length $T = 100$.]{\includegraphics[width=0.5\textwidth]{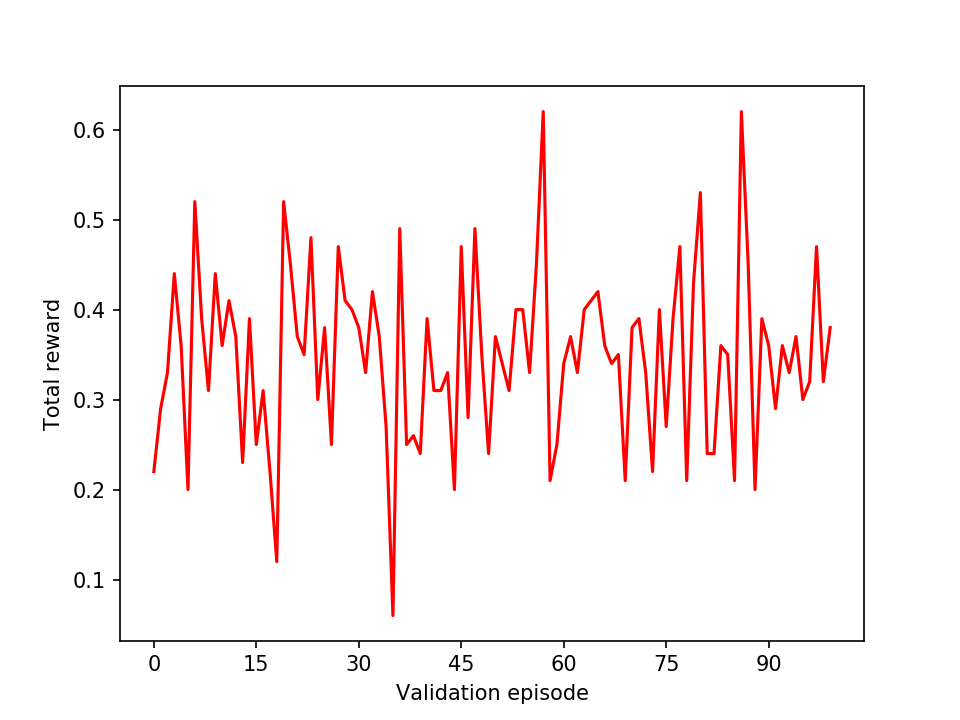}\label{fig:exp_rf_N5_random}}
  \caption{Average total rewards of a random agent on BF and RF for medium sequences.}
  \label{fig:medium_exp_random}
\end{figure}

\begin{figure}[!tbp]
  \centering
  \subfloat[$\rho_{BF}$ with $B = 400$, episodes length $T = 200$.]{\includegraphics[width=0.5\textwidth]{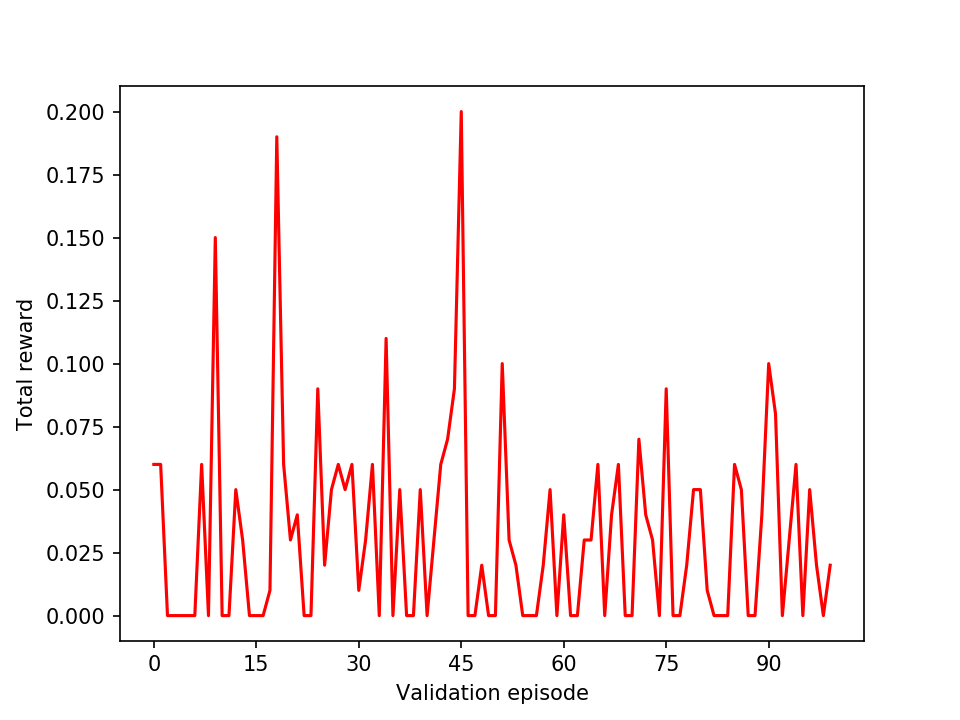}\label{fig:exp_bf_B400_random}}
  \hfill
  \subfloat[$\rho_{RF}$ with $N = 10$, episodes length $T = 100$.]{\includegraphics[width=0.5\textwidth]{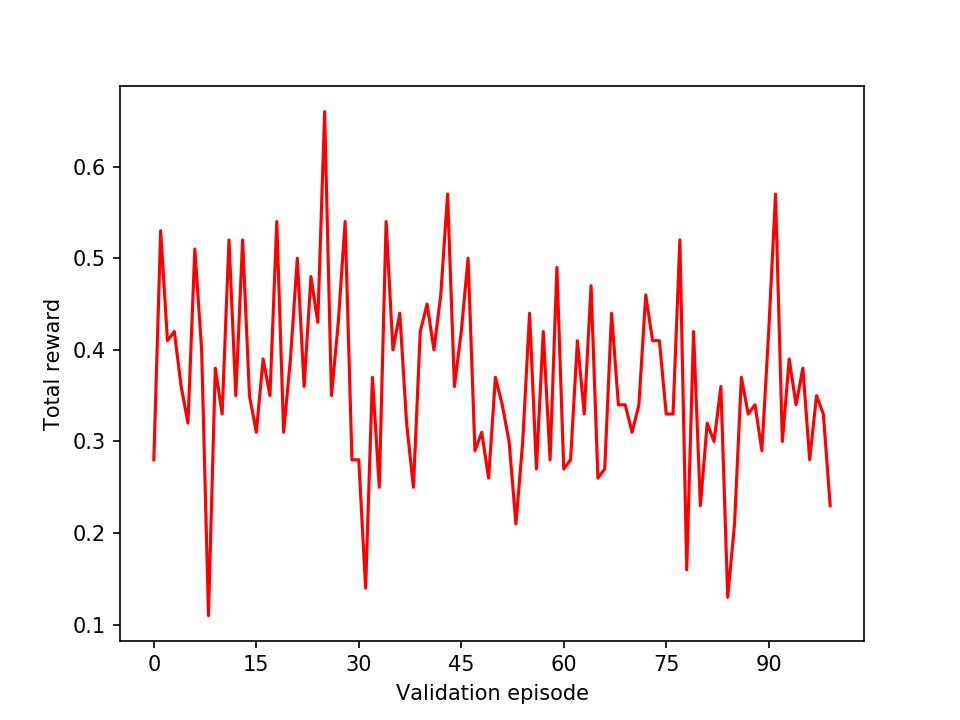}\label{fig:exp_rf_N10_random}}
  \caption{Average total rewards of a random agent on BF and RF for long sequences.}
  \label{fig:big_exp_random}
\end{figure}

Figures~\ref{fig:exp_bf_B80_random}, \ref{fig:exp_rf_N2_random}, \ref{fig:exp_bf_B200_random}, \ref{fig:exp_rf_N5_random}, and~\ref{fig:exp_bf_B400_random}, \ref{fig:exp_rf_N10_random} describe three comparisons between the random agents producing short, medium and long sequences respectively. As before, short means 80 bits in BF and 200 bits in RF, medium means 200 bits in BF and 500 bits in RF, long means 400 bits in BF and 1000 bits in RF. In each comparison, the average total rewards of $\rho_{RF}$ is higher than $\rho_{BF}$. From these results, we can assess that RF is a better formulation than BF for the PRNG task.

\begin{figure}[!tbp]
  \centering
  \subfloat[NIST score $0.4$]{\includegraphics[width=0.25\textwidth]{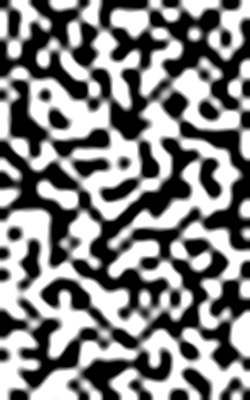}\label{fig:data_01}}
  \hfill
  \subfloat[NIST score $0.43$]{\includegraphics[width=0.25\textwidth]{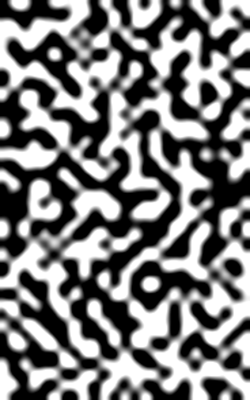}\label{fig:data_02}}
  \hfill
  \subfloat[NIST score $0.53$]{\includegraphics[width=0.25\textwidth]{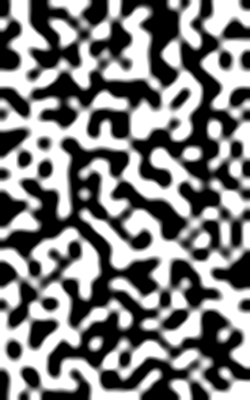}\label{fig:data_03}}
  \caption{A graphical representation of 3 sequences of 1000 bits generated by the same trained $\pi_{RF}$ with their NIST score. Images are obtained by stacking the 1000 bits in 40 rows and 25 columns, then ones are converted to $10\times 10$ white squares and zeros to $10\times 10$ black squares. The resulting image is smoothed.}
  \label{fig:data}
\end{figure}

Finally, figures~\ref{fig:data_01}, \ref{fig:data_02} and~\ref{fig:data_03} represent graphically three sequences of $1000$ bits generated by the same agent $\pi_{RF}$ after training, with NIST scores $0.4$, $0.43$ and $0.53$ respectively. The 1000 bits are stacked on 40 rows and 25 columns, then ones are converted to $10\times 10$ white squares and zeros to $10\times 10$ black squares, and the resulting image is smoothed.

\section{Conclusions and Future Work}\label{sec:conclusions}

In this paper we introduce a novel partially observable MDP modeling the task of generating PRNGs from scratch using DRL, denoted by RF for recurrent formulation. RF improves our previous MDP modeling in \cite{pasqualini2020pseudo} because it makes the action space size independent on the length of the generated sequence.

Experiments show that RF agents trained with PPO obtains at the same time a higher average NIST score and longer sequences, thus improving \cite{pasqualini2020pseudo} in two different ways. We use a PPO instance with a hidden state of an LSTM to encode significant features of the non observed portion of the sequence: as far as we know, this is an original idea for PRNG.

Experiments with a random agent show that RF is a better MDP modeling when compared with the binary formulation BF of \cite{pasqualini2020pseudo}, that is, when RF is compared with BF, obtains a higher average NIST score with longer sequences.

All this means that RF scales better to PRNG with longer periods.

However, while RF is a serious improvement over BF, the action space size grows as $2^N$, where $N$ is the bit-length of the appended sequence. Since DRL does not scale well to discrete and large action sets, see for instance \cite{zahavy2018learn}, this is a limitation for RF. In our experiments, we have found that $N > 10$ is not feasible for RF.
Moreover, the vanishing gradient is an obstacle to increasing the episodes length $T$ with recurrent neural networks, as shown in \cite{bengio1994learning}. In our experiments we have been unable to train RF with $T = 200$, so we consider $T = 100$ an upper bound for RF with PPO and the LSTM architecture described in Section~\ref{sec:nn}.
Since the PRNG period is $T\cdot N$, we can say that RF with all the hyperparameters described in this paper does not scale well over $1000$ bits.

Another problem of this approach is the sparse reward, which in general makes difficult for a RL agent to be trained.

The above remarks takes to the following list of possible future improvements/research topics:
\begin{itemize}
    \item Devise a different, less sparse, reward function.
    \item Devise an algorithm using RF modified with a continuous action space. In this way we can append sequences of any length without increasing the size of $\mathcal{A}$. Some preliminary experiments show that this approach could work but at the time of writing it still presents some performance issues. Work in progress.
    \item RF could be used to intelligently stack generated periods. For example, one could train a policy to stack sequences generated by one or multiple PRNGs, even with different sequence lengths. This would generate PRNGs with very long periods without a sensible drop in quality. This approach could use a mixture of state-of-the-art PRNGs, DRL generators like the ones seen in this paper, and so on. A challenge of this approach is how to measure the change in the NIST score when appending a random sequence to another.
    \item The upper bound given by the episodes length $T$ could be overcome, or at least mitigated, by other architectures capable of maintaining a memory for a time longer than recurrent neural networks, like attention models.
\end{itemize}

\clearpage

{}

{}

{}

\clearpage

\printbibliography

\end{document}